\documentclass[pra, reprint, amssymb, aps, superscriptaddress, floatfix, nofootinbib, citeautoscript, longbibliography]{revtex4-2}
\usepackage[utf8]{inputenc}
\usepackage{amsmath}  
\usepackage{amsfonts} 
\usepackage{graphicx} 
\usepackage{physics}
\usepackage{url}
\usepackage{xcolor}
\usepackage[T1]{fontenc}
\usepackage[colorlinks, linkcolor=blue, citecolor=blue, urlcolor=blue, breaklinks=true]{hyperref}
\usepackage{float}

\begin{document}

\title{\textbf{Correlating space, wavelength, and polarization of light: Spatio-Spectral Vector Beams}}

\author{Lea Kopf,$^{1}$ Rafael Barros,$^1$ and Robert Fickler$^1$\\
$^1$ \textit{Tampere University, Photonics Laboratory, Physics Unit, Tampere, FI-33720, Finland}}
\noaffiliation

\begin{abstract} 
Increasing the complexity of a light field through the advanced manipulation of its degrees of freedom (DoF) provides new opportunities for fundamental studies and technologies. 
Correlating polarization with the light's spatial or spectral shape results in so-called spatial or spectral vector beams that are fully polarized and have a spatially or spectrally varying polarization structure. 
Here, we extend the general idea of vector beams by combining both approaches and structuring a novel state of light in three non-separable DoF’s, i.e. space, wavelength, and polarization. 
We study in detail their complex polarization structure, show that the degree of polarization of the field is only unveiled when the field is narrowly defined in space and wavelength, and demonstrate the analogy to the loss of coherence in non-separable quantum systems.
Such light fields allow fundamental studies on the non-separable nature of a classical light field and new technological opportunities, e.g. through applications in imaging or spectroscopy.
\end{abstract}

\maketitle

\section{Introduction}
Increasing the complexity of a light field and the control of different degrees of freedom (DoF's) is beneficial for advancing research and technology. 
Increasingly complex structures realized by combining several DoF's have been studied in a myriad of experiments over the last decades and the enhanced understanding of the interplay of the DoF's has already enabled novel photonic technologies \cite{Rubinsztein_2017,forbes2021structured,piccardo2021roadmap,he2022towards,shen2022roadmap, Zhensong2023}.
Initially, many experiments have studied structuring transverse light fields as well as shaping the temporal profile of pulses, both in its scalar forms, i.e. with a uniform polarization structure \cite{weiner2000femtosecond, dickey2018laser}. 
The complexity of the light field's structure was further increased by including the polarization domain leading to beams with spatially non-uniform polarization distributions, i.e. spatial vector beams, as well as a temporally varying polarization vectors across the pulse duration.
Over the last years, this approach has been extended to combine all DoF's, e.g. the study of advanced spatio-temporal pulses of vectorial light fields \cite{chen2022synthesizing}, or the adaptation of established concepts such as complex transverse spatial fields to the spatio-temporal domain \cite{PhysRevX.6.031037,cao2022vectorial}.
Interestingly, the focus in most of the research efforts has been to generate complex polarization and spatial patterns over the temporal domain of light with much less attention to studying structured light fields in the time's complementary DoF, i.e. the spectral domain. 
Correlating polarization with the wavelength of a light pulse, for example, can be used for advanced sensing and pulse characterization schemes \cite{sano1980simple,del2003introduction,aspnes2014spectroscopic,Kopf_21,Jolly_21}.

\begin{figure}[b] 
    \centering
    \includegraphics[width = 0.45\textwidth]{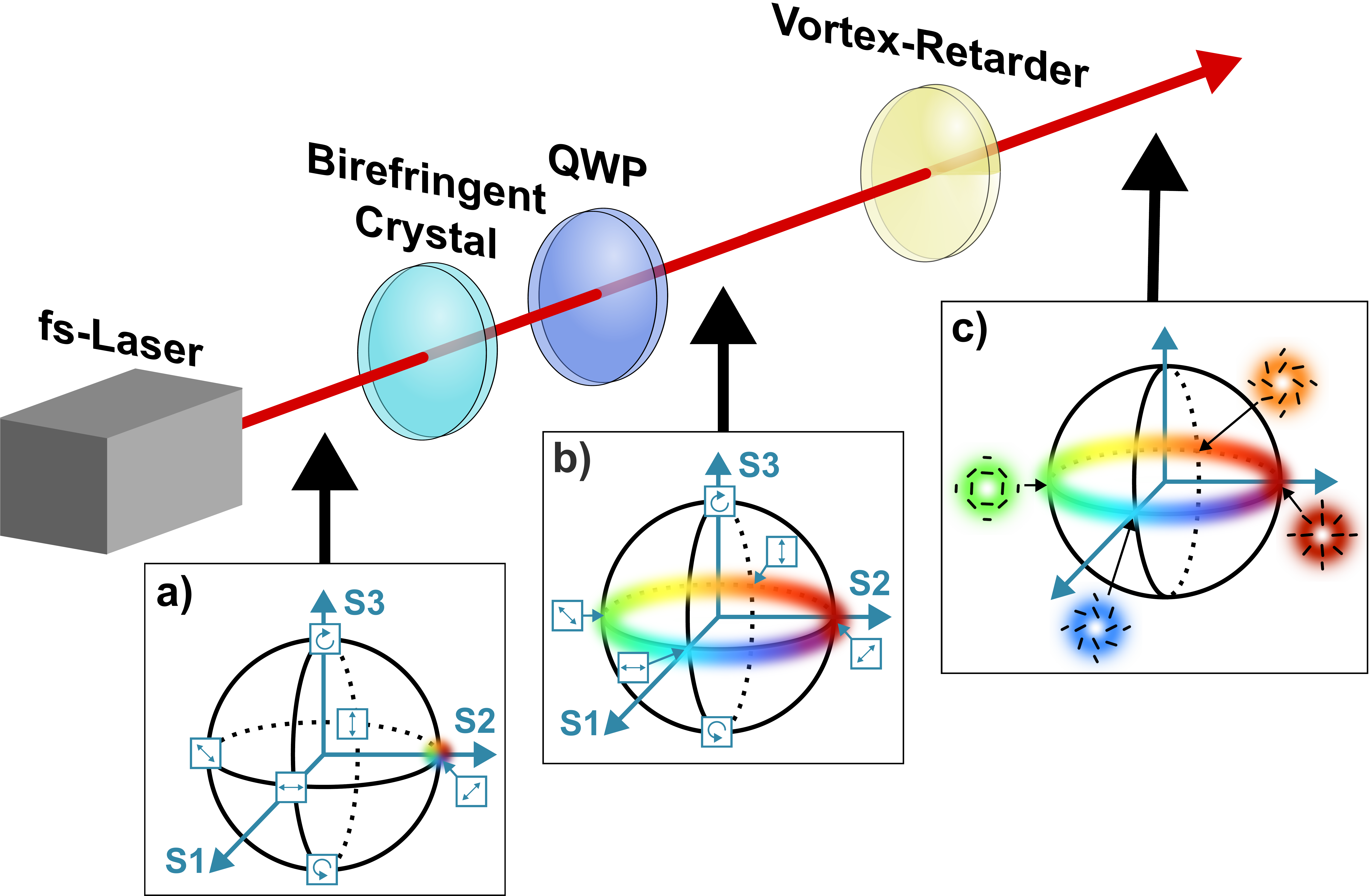}
    \caption{Conceptual idea. 
    A simple generation scheme consisting only of a birefringent crystal, a quarter wave-plate (QWP), and a vortex-retarder generates a complex light field where polarization, wavelength, and space are correlated. 
    The color in the insets symbolizes the different wavelength components of the light field.
    a) Poincaré sphere displaying that the light field is diagonally polarized. 
    b) The Poincaré sphere of the light field shows that each wavelength component has a different linear polarization. 
    c) The higher-order Poincaré sphere \cite{holleczek2010poincare,Milione2011} illustrates that each wavelength has a different spatial polarization pattern.}
    \label{fig:1}
\end{figure}
Here, we extend the idea of spatial and spectral vector beams by combining all three DoF's, namely polarization, space, and wavelength of the light field.
We term such light pulses spatio-spectral vector beams (SSVB's), which are light fields with a varying polarization structure in space as well as wavelength as shown in Fig.~\ref{fig:1}\,a). 
Using a simple setup consisting of only three optical elements placed along a single beam line, we are able to generate highly complex pulses of light for which any given wavelength (or transverse angular position) shows a different spatial (or spectral) polarization pattern.
We further show that all three DoF's are non-separable and the complex vectorial nature of the light field can only be observed when all DoF's are resolved.
By integrating over the light's transverse spatial extend or wavelength spectrum (or both), the beam would be seemingly unpolarized.
Finally, we detail the analogy of such complex structured light fields to a tripartite Greenberger–Horne–Zeilinger (GHZ) state, which describes three entangled particles in quantum optics.
Here, the DoF's of the classical light field act as the different quantum states in the GHZ description. 
We expect that the complexity of SSVB's combined with the ease of our generation technique will trigger novel studies on structured light fields and advance their applications in novel spectroscopy, imaging, or sensing technologies.

\section{Method}

We consider a SSVB of the form
\begin{equation}
        \vec{E}(\vec{r},\lambda)= \sqrt{\frac{I_0}{2}}\left[S_1(\vec{r}) F_1(\lambda) \hat{\epsilon}_1+S_2(\vec{r}) F_2(\lambda) \hat{\epsilon}_2\right]\,,
        \label{Eq_Field}
\end{equation}
where $\vec{r}$ is the position vector, $\lambda$ is the wavelength, $I_0$ is the total field intensity, $S_{1,2}$ and $F_{1,2}$ are normalized spatial and spectral basis functions, respectively, and $\hat{\epsilon}_{1,2}$ are unit polarization vectors.
We assume orthogonal polarization vectors with $\hat{\epsilon}_{i}\cdot\hat{\epsilon}_{j}=\delta_{ij}$, but allow for the spatial and spectral functions to have non-vanishing overlap. 
Thus, the function can describe a light field with non-separable polarization, space, and wavelength for zero overlap.

As a measure of the degree of tripartite non-separability of field \eqref{Eq_Field}, we consider the degree of polarization (DoP) of the light field. As we detail in appendix~\ref{appendix1}, the DOP is given by
\begin{equation}
    D=\sqrt{\frac{\left(\Lambda^S_{11}\Lambda^F_{11}-\Lambda^S_{22}\Lambda^F_{22}\right)^2+4|\Lambda^S_{12}|^2|\Lambda^F_{12}|^2}{\left(\Lambda^S_{11}\Lambda^F_{11}+\Lambda^S_{22}\Lambda^F_{22}\right)^2}}\,,
    \label{EQ_DegreeOfPolarization}
\end{equation}
where $\Lambda^S_{ij}=\langle S_i,S_j\rangle_{\Omega_S}$ and $\Lambda^F_{ij}=\langle F_i,F_j\rangle_{\Omega_F}$ ($j=1,2$) are the inner products between the spatial and spectral basis functions over finite regions $\Omega_S$ and $\Omega_F$ in space and wavelength, respectively. Experimentally, the DOP can be retrieved from the Stokes parameters resulting from intensity measurements integrated over $\Omega_S$ and $\Omega_F$.

Note that when integrating over the complete spatial and spectral DoF's ($\Omega_S=\mathbb{R}^2$ and $\Omega_F=\mathbb{R}$), Eq.~\eqref{EQ_DegreeOfPolarization} yields $D=|\Lambda^S_{12}\Lambda^F_{12}|$. 
Thus, when either the spatial or spectral basis functions are orthogonal, the measured light field is seemingly unpolarized.
The same argument holds for space or wavelength-only measurements, which yield reduced degrees of spatial and spectral coherence, respectively. 
Only joint measurements of the three DoF's unveil the full coherence of the input field, which is a signature of tripartite non-separability.
The apparent decoherence of the field \eqref{Eq_Field} upon partial tracing of one DoF is a by-product of its mathematical similarity to the GHZ states \cite{bouwmeester1999observation, Pan_00,Greenberger1989}.
Classically, such a tripartite non-separable behaviour can also be explained in an intuitive way. 
At any point in space where both spatial basis functions are non-vanishing, the polarization is spectrally non-uniform, i.e. a spectral vector beam \cite{Kopf_21}.
Similarly, at any wavelength present in both spectral basis functions, the polarization is spatially non-uniform, i.e. a spatial vector beam \cite{Zhan:09}.
Lastly, measuring the linear polarization state of the light field leads to a strong correlation between spatial and spectral structures.
In other words, specifying one DoF can project the remaining two in a bipartite non-separable field, which we will explore in more detail in the results section. 
We note that a related behaviour has been observed using multiple beams, each with a spatially varying polarization \cite{Balthazar:16}.

\subsection{Experimental Generation}

To create an SSVB, we use the process shown in Fig.~\ref{fig:1}.
First, Fourier-limited laser pulses with a duration of $\tau=220$\,fs and centered at a wavelength of 780\,nm are linearly polarized. 
The polarization is chosen to be diagonal with respect to the fast and slow axes of a 2\,mm long birefringent BaB$_2$O$_4$ (BBO) crystal with the optical axis oriented at $23.4^\circ$ from the propagation direction.
The propagation through the birefringent crystal coherently splits each pulse into two trailing pulses with equal intensities, linearly polarized along the crystal's fast ($f$) and slow ($s$) axes, producing a spectral vector beam \cite{Kopf_21}. The field at this stage is 
\begin{align}
    \vec{E}(\vec{r},\lambda)=& \sqrt{\frac{I_0}{2}}S_0(\vec{r})F_0(\lambda)\left[e^{i\pi c \delta t/\lambda}\hat{e}_f+ \right. \nonumber\\
    &\left. e^{-i\pi c \delta t/\lambda}\hat{e}_s\right]\,,
\label{EQ_ExpFieldPart}
\end{align}
where $S_0(\vec{r})$ is a Gaussian transverse spatial profile, $F_0(\lambda)$ is the spectral function, $c$ is the speed of light, and $\delta t$ is the birefringent time delay.
A subsequent quarter-wave plate oriented at 45$^\circ$ with respect to the fast axis of the crystal transforms the field's polarization components $\hat{e}_f$ and $\hat{e}_s$ to left and right circular polarizations, i.e. $\hat{e}_R$ and $\hat{e}_L$ respectively.
We set the time delay to $\delta t\approx \tau$ by choosing the appropriate crystal length and fine-tuning the crystal rotation, which ensures that the resulting linear polarization state in the wavelength domain rotates approximately once within one spectral bandwidth.
This beam is non-separable in wavelength and polarization, but still has only a single transverse structure, i.e. a Gaussian mode.
 
Next, we correlate the spatial DoF with polarization using an m=1 zero-order vortex half-wave retarder.
The vortex retarder has a constant half-wave retardance across the clear aperture with the fast axis oriented along $\theta=\phi/2$, where $\phi$ is the azimuthal angle. 
Upon transmission, the left and right circular polarization components change their handedness and acquire an azimuthally varying phase from 0 to 2$\pi$ \cite{Biener:02} resulting in
\begin{align}
    \vec{E}(\vec{r},\lambda)=& \sqrt{\frac{I_0}{2}}S_0(\vec{r})F_0(\lambda)\left[e^{i\phi}e^{i\pi \tau c/\lambda}\hat{e}_L+ \right. \nonumber \\
    &\left. e^{-i\phi}e^{-i\pi \tau c/\lambda}\hat{e}_R\right]\,.
    \label{Eq_ExperimentalField}
\end{align}
\begin{figure*}[ht] 
    \centering
    \includegraphics[width = \textwidth]{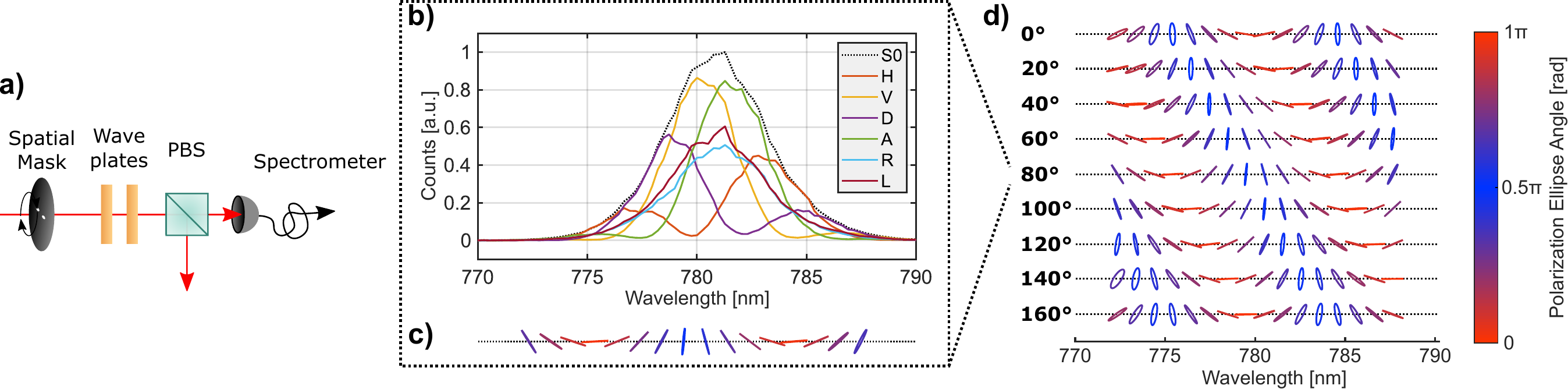}
    \caption{{Spectrally resolved measurements of the generated SSVB's.} a) The measurement setup filters space by using a spatial mask. Subsequently, it filters polarization using a half and quarter wave plate followed by a polarizing beam splitter (PBS). The final data acquisition is done in the spectral domain using a spectrometer. b) Polarization-resolved spectral measurements with a spatial mask rotation angle of 80$^\circ$. c) Reconstructed polarization ellipses over wavelength. The colors refer to the polarization ellipse rotation angle and are included for better visibility of the changes in polarization. d) The SSVB at different spatial mask rotation angles.}
    \label{fig:2}
\end{figure*}
 
The field \eqref{Eq_ExperimentalField} is exactly of the intended form of the SSVB \eqref{Eq_Field}, provided that we identify $S_{1,2}(\vec{r})=S_0(\vec{r})e^{\pm \phi}$ and $F_{1,2}(\lambda)=F_0(\lambda)e^{\pm i\pi\tau c/\lambda}$.
Note that while the spatial basis functions are mutually orthogonal ($\langle S_1,S_2\rangle_{\mathbb{R}^2}$=0), the overlap between the spectral basis functions is non-vanishing ($\langle F_1,F_2\rangle_{\mathbb{R}}\sim 1/e^2$).
The spectral overlap limits the quality of our SSVB, and is the cost for the simple method to produce it.
We note that the aforementioned limitation can be overcome by more advanced pulse shaping techniques to generate orthogonal temporal or spectral modes \cite{weiner2000femtosecond,monmayrant2010newcomer}, however, they lead to the same arguments as presented here.

\section{Results and Discussion}
\label{sec:Results}
\subsection{Characterization}

To characterize the SSVB \eqref{Eq_ExperimentalField}, we perform both individual and joint measurements of its polarization, wavelength, and spatial profile.
In the first set of measurements, we use the experimental setup depicted in Fig.~\ref{fig:2}\,a).
We send the SSVB through a spatial mask comprising two diametrically opposite 0.2\,mm wide apertures centred at the azimuthal angles $\phi_p$ and $\phi_p+\pi$, which roughly corresponds to two 17$^\circ$ wide slits and projects the field onto approximately $\vec{E}(\phi_p,\lambda)$.

We characterize the resulting field with spectrally-resolved polarization tomography, where the Stokes parameters are measured with a spectrometer, as shown for $\phi_p=80^\circ$ in Fig.~\ref{fig:2}\,b) with the corresponding wavelength-dependent polarization ellipses in c).
In Fig.~\ref{fig:2}\,d), we show different rotation angles of the spatial mask, which display a linear shift of the spectral-polarization structure. 
Note that the measured polarization states shown in Fig.~\ref{fig:2}\,c-d) are not strictly linear, as expected from \eqref{Eq_ExperimentalField}, but consist of elongated ellipses.
This results from the non-zero overlap of the spectral basis functions in our scheme and experimental imperfections such as the finite area of the spatial slits.

\begin{figure*}[ht] 
    \centering
    \includegraphics[width = \textwidth]{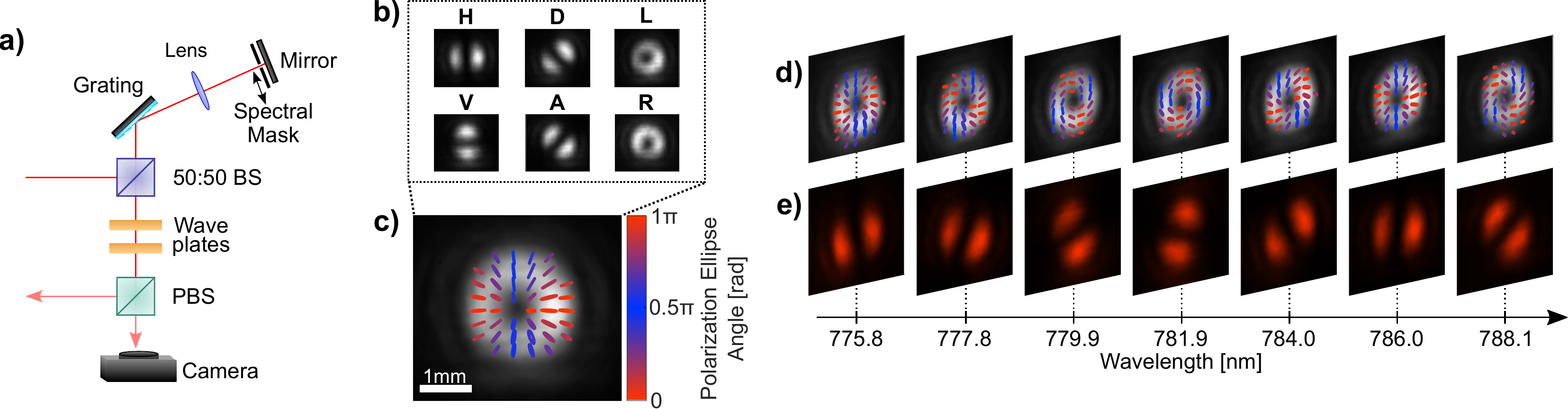}
    \caption{{Spatially resolved measurements of the generated SSVB.} a) The measurement setup filters wavelength and polarization using a monochromator, wave plates, and a polarizing beam splitter (PBS). The data is acquired in the spatial domain with a camera. b) The polarization measurements at a wavelength filter centred at 785.5\,nm. c) The reconstructed SSVB from the polarization measurements shown in b). The color corresponds to the rotation angle of the polarization ellipses and acts as a guide to the eye. d) The SSVB's at varying wavelength filter positions. e) The projection of the vector beam on the horizontal polarization at varying wavelength filter positions.} 
    \label{fig:3}
\end{figure*}

In the second set of measurements, we project the SSVB onto wavelength instead of space, using the setup depicted in Fig.~\ref{fig:3}\,a). 
A monochromator consisting of a diffraction grating, a mirror, and a tunable rectangular slit selects a central wavelength $\lambda_p$, which projects the field onto approximately $\vec{E}(\vec{r},\lambda_p)$. 
The projection resolution is fixed by the minimum bandwidth of around 2.1\,nm of the monochromator.
We characterize the spectrally filtered field through spatially-resolved polarization tomography, in which the Stokes parameters are measured with a CMOS camera. 
In Fig.~\ref{fig:3}\,b), we show the polarization-resolved images obtained at $\lambda_p=785.5\,$nm with the corresponding spatial vector beam in c).  
The spatial vector beams retrieved for different wavelengths are shown in Fig.~\ref{fig:3}\,d) and, to help visualize the wavelength-dependence of the retrieved field structures, the related images obtained at a fixed horizontal polarization are additionally shown in e).
The polarized images show a two-lobe structure that rotates across the spectrum, as expected from Eq.~\eqref{Eq_Field}.

\begin{figure}[t] 
    \centering
    \includegraphics[width = 0.35\textwidth]{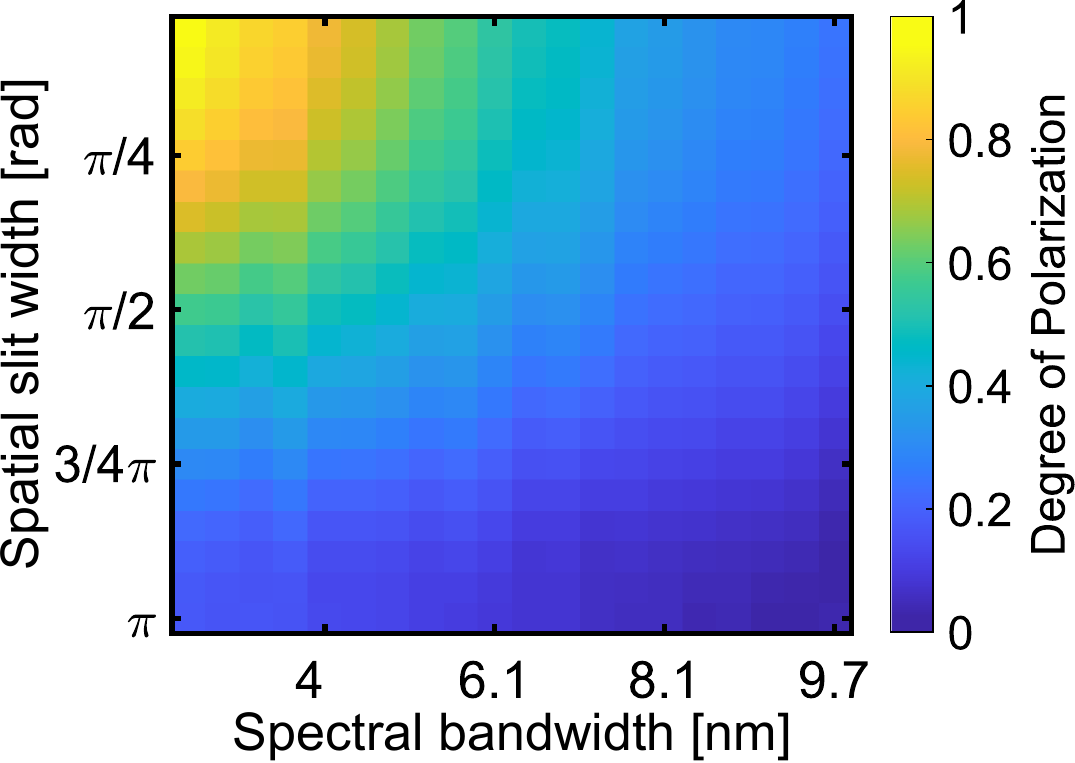}
    \caption{{The DoP as a function of the spatial and spectral filter bandwidths.}
    As both filters are enlarged, the DoP decreases due to the non-separability of space, wavelength, and polarization.}
    \label{fig:4}
\end{figure}

We further study that partially resolving the spatial and spectral domains can be used to control the DoP of the laser beam. 
To this end, we perform polarization measurements for different spatial and spectral bandwidths. 
We repeat the measurements shown in Fig.~\ref{fig:3}\,a) with different spectral resolutions at $\lambda_p=779.75$\,nm by changing the width of the rectangular slit in the monochromator. 
Then, a computational spatial mask with varying angular slit widths is added to the images in post-processing, simulating the physical mask used in Fig.~\ref{fig:2}\,a). 
From the measured Stokes parameters for each setting, we retrieve the DoP shown in Fig.~\ref{fig:4}.
By increasing the bandwidth of the wavelength filter or the slit width of the spatial mask, thereby averaging over the two DoF's, the beam's DoP decreases substantially. 
The initial DoP of 0.95, measured with a wavelength filter bandwidth of 2.1\,nm and a spatial slit mask 0.16\,rad wide, drops down to 0.04 when both wavelength and spatial filters are completely removed, showing excellent agreement with the simulations (see Fig.~\ref{fig:DoP} in the appendix).
Without resolving all DoF's, we thus have a seemingly unpolarized light field. 
Moreover, note that the minimum DoP obtained for spatially-resolved measurements is non-negligible ($\sim 0.2$), which agrees with the estimated overlap between the spectral basis functions of $1/e^2$. 

\subsection{Analogy with quantum mechanics}
As has been already recognized, classical non-separable light fields share a mathematical similarity with entangled quantum states \cite{Spreeuw_00,Aiello2015Quantumlike,Shen2021Creation}. 
In contrast to manifold discussions of two non-separable DoF's of light and the analogy to Bell states \cite{Karimi2015,Töppel2014}, SSVB's are mathematically similar to the description of three-particle GHZ states. 
The GHZ state has the form
\begin{equation}
    \ket{\Psi}=\frac{1}{\sqrt{2}}\left(\ket{000}+\ket{111}\right)\,,
    \label{EQ_GHZ}
\end{equation}
where $\ket{0}$ and $\ket{1}$ denote the two possible states of each quantum system expressed in terms of qubits in the computational or $z$-basis.
The mutually unbiased bases to $z$ are the $x$- and $y$-bases, defined in the Dirac notation as $\ket{\pm x}=(\ket{0}\pm\ket{1})/\sqrt{2}$ and $\ket{\pm y}=(\ket{0}\pm i\ket{1})/\sqrt{2}$, respectively. 

To demonstrate the non-separability of the SSVB, we closely follow the GHZ argument \cite{Pan_00}, which focuses on the simultaneous measurements of all three qubits in the $x$-basis ($xxx$ measurement). 
In the original GHZ argument, the two possible measurement outcomes in each basis are ascribed with the values $+1$ and $-1$. 
Together with the assumption of a local realistic theory, it is possible to show that classical correlations require the product of the measurement outcomes in the $x$-basis to be $+1$, while for quantum correlations the outcome will be $-1$.
By adapting these ideas to SSVB's and showing a similar behaviour, we can use the GHZ argumentation to demonstrate the non-separability of the classical light field.

We assign the notation \eqref{EQ_GHZ} to the SSVB by representing the spatial, spectral and polarization modes as the first, second, and third entries of the state vector, respectively.
For the spatial DoF, the $z$ basis is formed by the spatial functions $S_{1,2}(\vec{r})=S_0(\vec{r})e^{\pm i \phi}$, which contains counter-rotating spiral wavefronts carrying opposite orbital angular momenta \cite{PhysRevA.45.8185}, and indistinguishable intensity profiles.
On the other hand, the $x$ ($y$) basis modes have two-lobe structures oriented along $0^\circ$ and $90^\circ$ ($45^\circ$ and $-45^\circ$) \cite{padgett1999poincare}, and can be distinguished by narrow slits oriented along these directions \cite{fickler2012quantum, plick2015violation}. 
We implement these slits by using the spatial mask already presented in the measurements of Fig.~\ref{fig:2}.

In the wavelength domain, the $z$ basis is formed by the spectral functions $F_{1,2}(\lambda)=F_0(\lambda)e^{\pm i\pi\tau c/\lambda}$, which cannot be distinguished by measurements of spectral amplitude. 
On the other hand, the modes composing the $x$ basis are $F_{1x}\propto F_0(\lambda)\cos(\pi\tau c/\lambda)$ and $F_{2x}\propto F_0(\lambda)\cos(\pi\tau c/\lambda-\pi/2)$, while those of the $y$ basis are $F_{1y}\propto F_0(\lambda)\cos(\pi\tau c/\lambda-\pi/4)$ and $F_{2y}\propto F_0(\lambda)\cos(\pi\tau c/\lambda-3\pi/4)$. 
In analogy to the spatial domain, we measure in the $x$ and $y$ bases by choosing, in post-processing, four wavelengths spaced by $1/\tau$ at which the intensity difference between the horizontal/vertical or diagonal/anti-diagonal polarization components are maximized.

Lastly, for the polarization DoF, we define the $z$ basis vectors as the left and right circularly polarized components of the light field.
Consequently, the $x$ basis corresponds to horizontal/vertical linear polarizations, and the $y$ basis to diagonal/anti-diagonal linear polarizations. 
A graphical overview of the chosen bases is given in Fig.~\ref{fig:Bases} in the appendix.

\begin{figure}[ht!] 
    \centering
    \includegraphics[width = 0.325\textwidth]{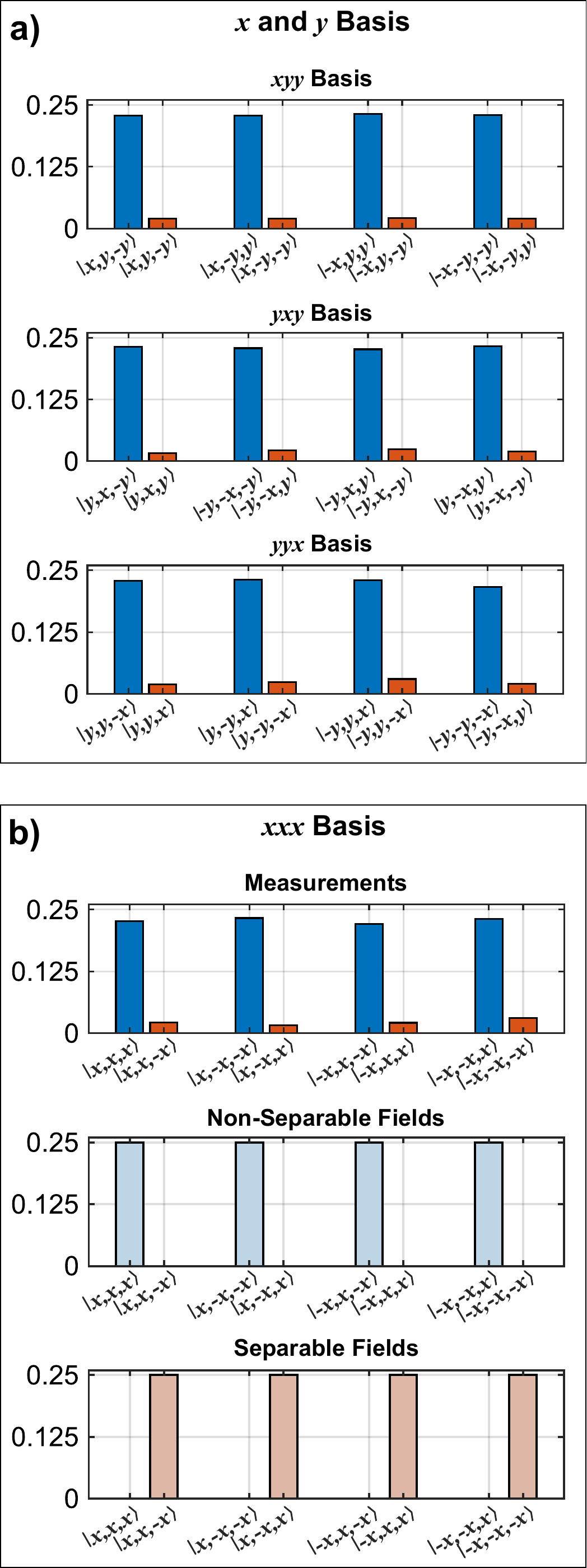}
    \caption{{Measurements to show the non-separability of the SSVB.} a) Measurement probabilities where one DoF is measured in the $x$-basis and the other two are measured in the $y$-basis.
    The blue bars correspond to the expected states in the mathematical description of a GHZ state, the red bars correspond to nonexistent states.
    b) Measured detection probabilities when all DoF's are measured in the $x$-basis.
    The measurements in the top row nicely show, that the description of the non-separable state (middle row) is dominant over the expected result for separable fields (lower row).} 
    \label{fig:5}
\end{figure}

Fig.~\ref{fig:5} shows the measurement results for all terms of the normalized state \eqref{EQ_GHZ} written in the $x$ and $y$ bases. 
More information on the normalization procedure can be found in the appendix.
Clearly, the $xxx$ measurements show that the product of the three outcomes mostly yields $+1$, with a small contribution of $-1$ due to experimental limitations and imperfections. 
We regard this result as an interesting way to show the non-separability between polarization, spatial profile and wavelength by exploring the mathematical isomorphism with the GHZ state.

\section{Conclusion} 
In this article, we introduced and experimentally realized a light field with a complex polarization structure in space as well as wavelength - a spatio-spectral vector beam. 
Our measurements show that the SSVB is well defined in its three DoF's if, and only if, they are measured jointly, while individual measurements result in apparent incoherence, i.e. a reduced degree of polarization, similar to the loss of coherence in non-separable quantum systems.
In the future, it will be interesting to extend the concept further to more complex spatio-spectral patterns, e.g. by including spatial Poincaré beams \cite{Beckley:10} or studying SSVB's in the context of optical skyrmions \cite{doi:10.1021/acsphotonics.1c01703}. 
Not only do SSVB's contribute to a better understanding of the complex interplay between different DoF's, they might also enable applications benefiting from the complex correlations and simple generation.
In contrast to earlier work, where spatial and spectral vector beams have been used for high-speed tracking of changes in space \cite{berg2015classically} and spectrum \cite{Kopf_21}, the presented light fields could allow a fruitful combination of the two leading to novel schemes for advanced sensing methods in all optical domains simultaneously.
In addition, our results might be applied in quantum optics experiments when complex hyper-entangled states are utilized \cite{barreiro2005generation, achatz2023simultaneous}.
Finally, it will be interesting to study the possible benefits of SSVB's in nonlinear light-matter interactions of structured light \cite{buono2022nonlinear}.

%

\newpage

\section*{Acknowledgements} 
    The authors acknowledge the support of the Academy of Finland through the Competitive Funding to Strengthen University Research Profiles (decision 301820) and the support of the Photonics Research and Innovation Flagship (PREIN - decision 320165). LK acknowledges the support of the Vilho, Yrjö and Kalle Väisälä Foundation of the Finnish Academy of Science and Letters. RB acknowledges the support of the Academy of Finland through the postdoctoral researcher funding (decision 349120). RF acknowledges the support of the Academy of Finland through the Academy Research Fellowship (decision 332399).

\clearpage

\appendix
\renewcommand\thefigure{\thesection\arabic{figure}}    
\setcounter{figure}{0} 

\section{Simulation of the degree of polarization}
\label{appendix1}

The degree of polarization (DoP) of a field $\vec{E}(\vec{r},\lambda)$ is defined as \cite{Kagalwala_15} 
\begin{equation}
D=\frac{1}{\langle I_0 \rangle}\sqrt{S_H^2+S_D^2+S_R^2}
    \label{EQ_DoP_a}
\end{equation}
where $\langle I_0 \rangle $ is the total average intensity, and $S_j=2\langle I_j\rangle-\langle I_0 \rangle$ ($j=H,D,R$) are the so-called Stokes parameters. The indices H, D, and R denote polarization projections along the horizontal, diagonal, and right-circular polarizations, without loss of generality.
The average intensities $\langle I_j\rangle$ are given by 
\begin{equation}
    \langle I_j \rangle=\int_{\Omega_S} d^2\vec{r}\int_{\Omega_F} d\lambda\,|\vec{E}(\vec{r},\lambda)\cdot\hat{\epsilon_j}|^2,
\end{equation}
where $\Omega_S$ and $\Omega_F$ are the integration regions in the spatial and spectral domains, respectively. Using the field expressed in Eq. (1) of the main text, we obtain 
\begin{equation}
    \langle I_j \rangle = \frac{I_0}{2}\sum_{i=1,2} \Lambda^S_{ii} \Lambda^F_{ii} |\Lambda^P_{ij}|^2 +I_0\textrm{Re}\left( \\ \Lambda^S_{12} \Lambda^F_{12}\Lambda^P_{1j}\Lambda^{P*}_{2j}   \right)\,,
    \label{EQ_IntensityComponents_a}
\end{equation}
where $\Lambda^P_{ij}=\hat{\epsilon}_i\cdot\hat{\epsilon}_j^{\,*}\,$, and where
\begin{eqnarray}
    \Lambda^S_{ik}&=&\int_{\Omega_S} d^2 \vec{r}\, S_i(\vec{r})S_k^*(\vec{r})\,,\\
    \Lambda^F_{ik}&=&\int_{\Omega_F} d\lambda\, F_i(\lambda)F_k^*(\lambda)\,,
\end{eqnarray}
with $i,k=1,2$, are the overlaps of the spatial and spectral basis functions of $\vec{E}(\vec{r},\lambda)$ over $\Omega_S$ and $\Omega_F$.
Finally, by inserting the Stokes parameters calculated from \eqref{EQ_IntensityComponents_a} into \eqref{EQ_DoP_a}, we obtain 
\begin{equation}
    D=\sqrt{\frac{\left(\Lambda^S_{11}\Lambda^F_{11}-\Lambda^S_{22}\Lambda^F_{22}\right)^2+4|\Lambda^S_{12}|^2|\Lambda^F_{12}|^2}{\left(\Lambda^S_{11}\Lambda^F_{11}+\Lambda^S_{22}\Lambda^F_{22}\right)^2}}\,,
    \label{EQ_DegreeOfPolarization_a}
\end{equation}
which is the expression we use in the main text.

\begin{figure}[ht] 
    \centering
    \includegraphics[width = 0.45\textwidth]{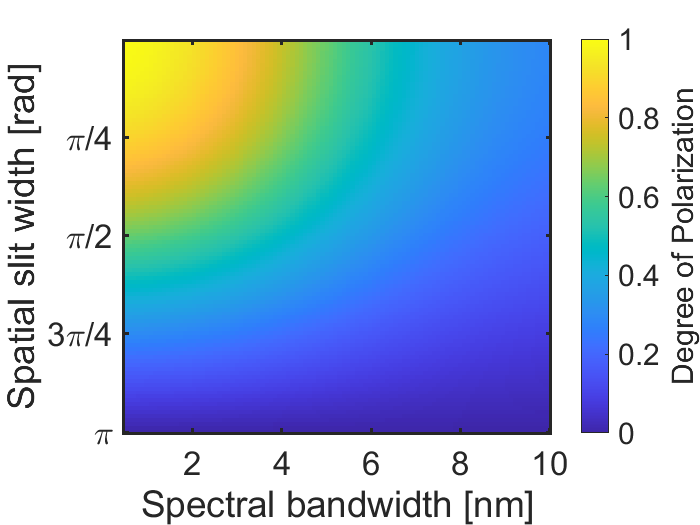}
    \caption{Simulated DoP for different spatial and spectral measurement bandwidths.} 
    \label{fig:DoP}
\end{figure}

In Fig.~\ref{fig:DoP}, we show the DoP calculated with Eq.~\eqref{EQ_DegreeOfPolarization_a} as a function of the spatial and spectral measurement bandwidths.
We use the spatial and spectral basis functions defined in Eq.~(4) of the main text, while considering a Gaussian spatial profile with approximately $w_0=600\,\mu$m beam waist and a Gaussian spectral profile corresponding to a pulse duration of $\tau=220$\,fs, matching the parameters used in the experiment.
The spatial bandwidth corresponds to the opening angle of an angular mask, while the spectral bandwidth corresponds to the width of a sharp window centred at $780$\,nm. As expected, the simulated DoP has excellent agreement with the measured DoP shown in Fig.~\ref{fig:4} of the main text.

\section{Analogy to the quantum description}
\label{appendix2}
\subsection{Bases}

\begin{figure*}[ht] 
    \centering
    \includegraphics[width = 0.75\textwidth]{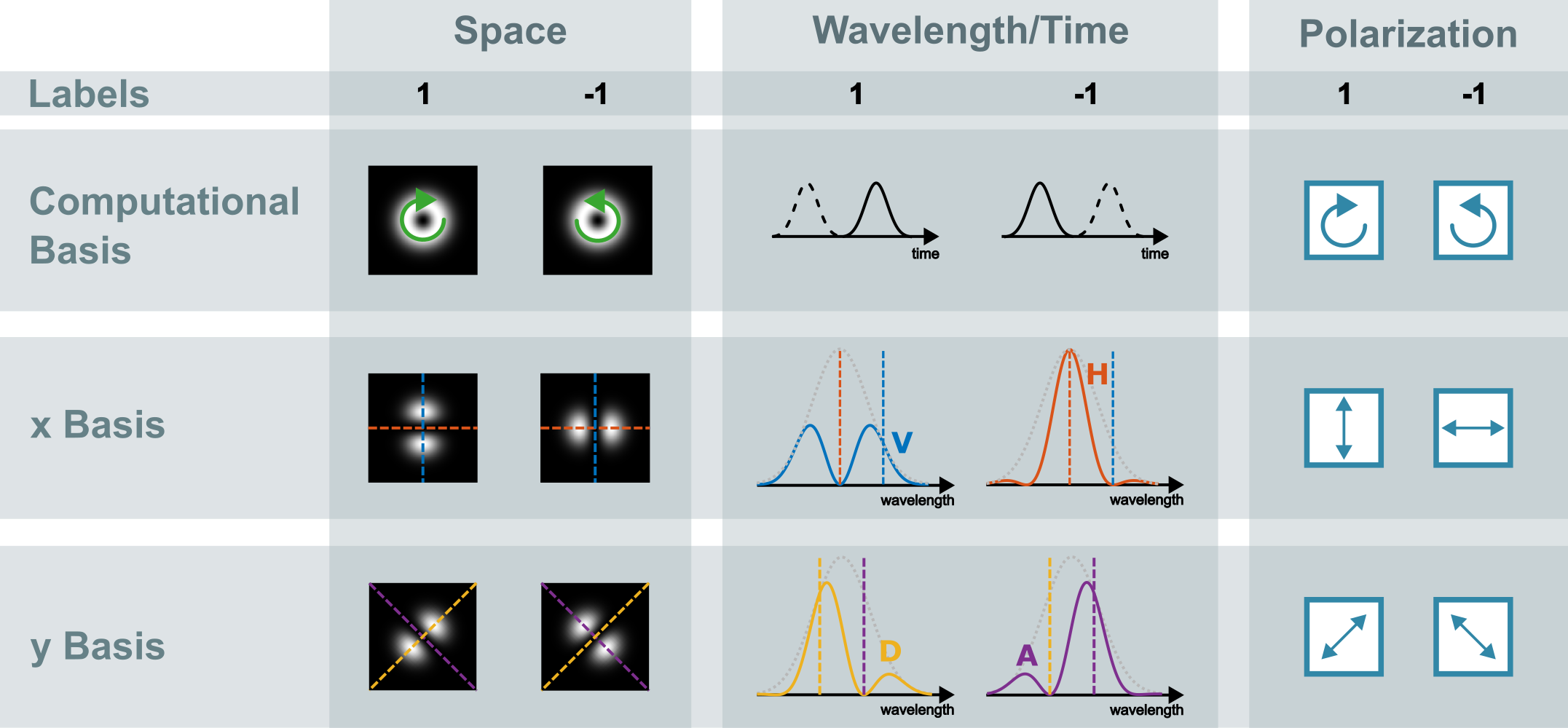}
    \caption{Overview of the chosen mutually unbiased bases and assigned labels in the three different DoF's. 
    The projections in the spatial and spectral domains on infinitesimal small slits are depicted by the dotted lines.}
    \label{fig:Bases}
\end{figure*}

Fig.~\ref{fig:Bases} shows an overview of the mutually unbiased bases and the assigned labels needed to show the non-separability in analogy to the quantum description.
The following list gives additional descriptions to the shown figure.
\begin{itemize}
    \item Space domain: The computational basis in the space domain corresponds to a light field with counter-rotating spiral phases as discussed in the main text. 
    Here, the basis states are two doughnut-shaped intensity structures with opposite azimuthal phase gradients as indicated by the green arrows in Fig.~\ref{fig:Bases}.
    The complementary $x$-bases correspond to a pair of two-lobbed structures which are rotated by 90$^\circ$ with respect to each other.
    The $y$-basis consists of two similar modes that are rotated by 45$^\circ$ with respect to the $x$-basis \cite{padgett1999poincare}.
    In the measurements, we approximate that a narrow spatial slit projects on a superposition of orthogonal spatial modes \cite{plick2015violation}.
    These slits are depicted by dotted lines in Fig.~\ref{fig:Bases}.

    \item Wavelength/Time domain: The computational basis for the spectral/temporal DoF is formed by two time-delayed pulses, which correspond to the functions $F_{1,2}(\lambda)=F_0(\lambda)e^{\pm i\pi\tau c/\lambda}$ given in the main text. 
    The $x$ and $y$ basis are made up of four specific frequencies at which the intensity difference between the horizontal/vertical or diagonal/anti-diagonal polarization components are maximized (shown in Fig.~\ref{fig:Bases} as dotted lines).
    We follow the same argumentation as in the spatial domain: The transmission through a very narrow slit, i.e. strongly filtering in the frequency domain, approximates the projection on a superposition of modes.
    
    \item Polarization: The computational domain is defined by the left and right circularly polarized components of the light field.
    The $x$ basis projects on horizontal/vertical linear polarization, while the $y$ basis projects on diagonal/anti-diagonal linear polarization.
\end{itemize}

\subsection{Normalization}
In the measurements, the intensity of the states projected onto polarization and space in the chosen bases are measured in the spectral domain.
In the Dirac notation all possible measurement outcomes in a basis should add up to a probability amplitude of 1.
Thus, the intensity measurements have to be normalized.
First, the measured intensity signal is normalized with respect to the intensity summed up over all polarizations, i.e. with respect to the $S_0$ Stokes parameter. 
Then, the measurements of all possible permutations of the basis are summed up to find a normalization constant $N$.
In the $xyy$-basis the state for example reads
\begin{align}
    \ket{\phi}=&\frac{1}{N} \left( \textcolor{blue}{\alpha}\ket{x,y,-y}+\textcolor{red}{\beta}\ket{x,y,y}+\textcolor{blue}{\gamma}\ket{x,-y,y}\right.\nonumber \\ 
    & \left. +\textcolor{red}{\delta}\ket{x,-y,-y}+    \textcolor{blue}{\epsilon}\ket{-x,y,y}+ \textcolor{red}{\zeta}\ket{-x,y,-y}+\right.\nonumber \\ & 
    \left. \textcolor{blue}{\eta}\ket{-x,-y,-y}+\textcolor{red}{\theta}\ket{-x,-y,y} \right)
\end{align}
where ideally for a non-separable state ${\textcolor{red}{\beta}}={\textcolor{red}{\delta}}={\textcolor{red}{\zeta}}={\textcolor{red}{\theta}}=0$. 
The normalization factor is given by
\begin{equation}
    N=\abs{\textcolor{blue}{\alpha}}^2+\abs{\textcolor{red}{\beta}}^2+\abs{\textcolor{blue}{\gamma}}^2+\abs{\textcolor{red}{\delta}}^2+\abs{\textcolor{blue}{\epsilon}}^2+\abs{\textcolor{red}{\zeta}}^2+\abs{\textcolor{blue}{\eta}}^2+\abs{\textcolor{red}{\theta}}^2.
\end{equation}

\end{document}